\documentclass{article}

\usepackage[latin1]{inputenc}
\usepackage{relsize}
\usepackage{amsmath,amsfonts,amssymb}
\usepackage{graphicx,color,psfrag,nicefrac,units}
\usepackage{achemso}

\newcommand{\dd}{\mathrm{d}}
\newcommand{\naber}{\nabla_{\Vektor{r}}}
\newcommand{\nabzr}{\nabla_{\Vektor{r}}^2}
\newcommand{\nabeR}{\nabla_{\Vektor{R}}}
\newcommand{\hrt}{h(\Vektor{r},t)}
\newcommand{\hrst}{h(\Vektor{r}',t)}
\newcommand{\hkt}{h(\Vektor{k},t)}
\newcommand{\hksts}{h(\Vektor{k}',t')}
\newcommand{\sbl}{\sigma\beta\ell^2}
\newcommand{\kb}{\kappa\beta}
\newcommand{\grR}{G(\Vektor{r}\!-\!\Vektor{R})}
\newcommand{\intlim}[3]{\!\int\limits_{#2}^{#3} {\! #1\,}}
\newcommand{\Onsager}[2]{\Lambda(\Vektor{#1}\!-\!\Vektor{#2}')}
\newcommand{\Onsagerk}[1]{\Lambda(#1)}
\newcommand{\Mittel}[1]{\left\langle #1\right\rangle}
\newcommand{\Exp}[1]{\exp\left\lbrace #1\right\rbrace}
\newcommand{\Vektor}[1]{\mathbf{#1}}
\newcommand{\stochteil}{\zeta}
\newcommand{\stochmemb}{\xi}
\newcommand{\stexyt}[1]{\stochteil_{#1}(t)}		
\newcommand{\smvrt}{\stochmemb(\Vektor{r},t)}
\newcommand{\smvkt}{\stochmemb(\Vektor{k},t)}

\begin{document}

\title{Curvature-coupling dependence of  membrane protein diffusion
  coefficients}

\author{Stefan M.~Leitenberger, Ellen Reister-Gottfried, and Udo
  Seifert\\\emph{\mbox{II}.~Institut f\"ur Theoretische Physik, Universit\"at
    Stuttgart, 70550 Stuttgart, Germany}}

\maketitle

\begin{abstract}
We consider the lateral diffusion of a protein interacting with the curvature
of the membrane. The interaction energy is minimized
if the particle is at a membrane position with a certain curvature that agrees
with the spontaneous curvature of the particle.  We employ stochastic
simulations that take into account both the thermal fluctuations of the
membrane and the diffusive behavior of the particle. In this study we neglect
the influence of the particle on the membrane dynamics, thus the membrane
dynamics agrees with that of a freely fluctuating membrane. Overall, we find
that this curvature-coupling substantially enhances the diffusion
coefficient. We compare the ratio of the projected or measured diffusion
coefficient and the free intramembrane diffusion coefficient, which is a
parameter of the simulations, with analytical results that rely on several
approximations. We find that the simulations always lead to a somewhat smaller
diffusion coefficient than our analytical approach. A detailed study of the
correlations of the forces acting on the particle indicates that the diffusing
inclusion tries to follow favorable positions on the membrane, such that
forces along the trajectory are on average smaller than they would be for
random particle positions.
\end{abstract}


\section{Introduction}
During the last decade it has become more and more apparent that lateral
diffusion of proteins in membranes plays a crucial role in cellular
functioning~\cite{Lippincott:2001,Marguet:2006}. Therefore, a whole range of
experimental techniques has been developed, which is constantly being improved
in order to determine accurate values of lateral protein diffusion
coefficients~\cite{Lommerse:2004,Chen:2006}. The most important methods include
fluorescence recovery after photo bleaching~\cite{Reits:2001,Lippincott:2003},
fluorescence correlation spectroscopy~\cite{kohl,Thompson:2002}, or single particle
tracking~\cite{saxton,Kusumi:2005}. While a large amount of data has been
collected with these techniques the interpretation of results always depends
on reliable models for the diffusive process. In some situations, like
restricted diffusion due to
corrals~\cite{Kusumi:2005,Tomishige:1998,Edidin:1991}, a certain, often rather
crude, qualitative explanation is easily found, in other situations, however,
this is by no means the case and a reliable quantitative interpretation can
only be achieved, if corresponding theoretical calculations or simulations are
performed.

Only recently an increased interest in lateral diffusion has emerged from a
theoretical viewpoint.  In order to compare theoretical results with
experiments it is necessary to take into account various aspects of the
particular system. These include the nature of the membrane the particle is
diffusing in, the properties of the diffusing particle, or the experimental
method with which diffusion coefficients are determined. A very important
aspect in both theoretical calculations and the analysis of experimental
results is that the membrane must not be regarded as a flat plane but is often
structured such that regions with higher and lower curvatures appear. For
example, this must be accounted for in the study of diffusion in the membranes
of the endoplasmic reticulum. Neglecting the influence of the membrane shape
leads to considerable errors in the determination of diffusion
coefficients~\cite{Sbalzarini:2006}. Several analytical and simulational
studies have been performed that regard diffusion on various fixed curved
surfaces~\cite{Aizenbud:1982,Holyst:1999,Plewczynski:2000,Faraudo:2002,King:2004,Yoshigaki:2007,Naji:2007}.

But even if a membrane appears to be flat on average, it is subject to thermal
fluctuations that lead to rapid shape changes around the flat
configuration~\cite{Brochard:1975,seifert}. Neglecting the influence of the
membrane on the movement of the particle these fluctuations that depend on
properties like bending rigidity, surface tension, proximity to substrates or
other membranes, etc., have an influence on the measured values for lateral
diffusion coefficients because experiments usually regard the path of the
inclusion projected on a flat reference plane instead of the actual path along
the membrane. Compared to the intramembrane diffusion coefficient the measured
diffusion coefficient will be the smaller the stronger the fluctuations. This
was initially pointed out by Gustafsson and Halle~\cite{Gustafsson:1997}; the
quantitative evaluation of this effect for free intramembrane diffusion was
performed more recently in independent work by us and two other groups using
analytical calculations~\cite{ellen_projection,Gov:2006} and
simulations~\cite{ellen_drift,Naji:2007}.

All studies mentioned in the last two paragraphs take into account the
influence of the shape of the membrane on measured diffusion coefficients, but
otherwise neglect any interaction between membrane and protein. Considering
that a protein also has certain physical properties it must be assumed that
interactions between the protein and the membrane exist that influence the
diffusion coefficient. This assumption is corroborated by experimental
findings: for example after photoactivation of bacteriorhodopsin (BR) in model
membranes the lateral diffusion coefficient is reduced by a factor of
five~\cite{kahya}. This reduction is attributed to oligomerization of BR upon
activation caused by structural changes that influence protein-lipid
interactions. There are many other experimental examples that indicate
that the interactions between membrane and protein have a significant
influence on lateral diffusion, see for example
refs.~\cite{Vereb:2003,Forstner:2006}.  An important property of a membrane
compared to a flat surface is the curvature. A variety of studies mainly using
particle based simulations are concerned with the influence of inclusions with
a certain intrinsic curvature on membrane shape and lateral
diffusion~\cite{Weikl:1998,Weikl:2001,Fournier:1999,Fournier:2003,Fosnaric:2006,Cooke:2006,Deserno:2007,Blood:2006,Ayton:2007}. In
earlier work we calculated effective diffusion coefficients for particles with
a bending rigidity and a spontaneous curvature~\cite{ellen_projection}. These
calculations revealed that the additional interaction that tries to move the
particle to positions on the membrane where the curvature agrees with the
particle's spontaneous curvature, leads to an increase in the diffusion
coefficient.

\begin{figure}
 \centering
 \includegraphics{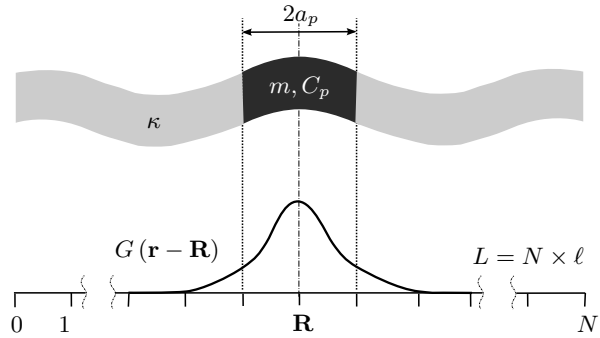}
 \caption{Sketch of the curvature-coupled model for an inclusion at position
   $\mathbf R$, with area $\pi a_p^2$, a bending rigidity $m$ and a
   spontaneous curvature $C_p$ in a model membrane with the bending rigidity
   $\kappa$ and the effective surface tension $\sigma$. The membrane is mapped
   to a two-dimensional $N\times N$ lattice with the lattice spacing
   $\ell$ of which only a one-dimensional cut is shown here. The lateral dimension of the system is $L=N\times\ell$. In the simulations,
   the area of the inclusion is determined by the weighting function $\grR$, which is
   set to be a Gaussian.}
 \label{fig:sketch1}
\end{figure}

Our previous work on curvature-coupled diffusion effectively describes
diffusion of a point-like particle and relies on several approximations
\cite{ellen_projection,ellen_drift}. One of these, the so-called pre-averaging
approximation, assumes that membrane fluctuations on all possible length
scales have a much shorter relaxation time than the time it takes the particle
to diffuse the corresponding distance. In this paper we introduce a scheme to
simulate the diffusion of a particle with a certain extension, a bending
rigidity and a spontaneous curvature in a thermally fluctuating membrane. A
sketch of the considered system with the relevant physical parameters is given
in fig.~\ref{fig:sketch1}.

The present method is no longer restricted to certain relative timescales of
diffusion and membrane fluctuations. The additional energy of the inclusion is
introduced by replacing a patch of membrane that is described by the Helfrich
Hamiltonian with a new Helfrich-like term with a different bending rigidity
and a spontaneous curvature. The extension of the particle is modeled by a
Gaussian weighting function in order to have a smooth crossover from the bare
membrane to the particle. Obviously the system is dominated by two dynamic
processes: the shape fluctuations of the membrane and the particle diffusion.
Two Langevin-equations, one for the membrane, the other for the particle, are
derived from the energy of the system. Our simulation scheme consists of the
numerical integration in time of these coupled equations. Apart from
performing simulations we also analytically evaluate the coupled dynamic
equations by use of a perturbation theory that neglects the influence of the
particle on the membrane and assumes that membrane relaxation times are much
smaller than corresponding diffusive time scales. In order to compare with
these analytical calculations and to reduce the computational effort, we
restrict our simulations in the current study such that the membrane dynamics
is also not influenced by the diffusing particle. The influence on membrane
movement will be considered in future work. The main quantity of interest is
the ratio of the curvature-coupled and the intramembrane diffusion coefficient
as a function of the membrane parameters bending rigidity and surface
tension. The latter coefficient is a parameter of our scheme and resembles the
free diffusion coefficient of the particle if no additional force were acting
on it. The application of both our approaches shows that curvature-coupling
leads to increased diffusion. However, the comparison of our analytic
calculations with the simulation results reveals a systematic difference. In
order to gain insight into the reason for these discrepancies we study force
correlation functions that are the main contributions to the diffusion
constant.

The paper is organized as follows: In the following section we explain the
model for the membrane dynamics and the Langevin-equation for the inclusion.
In sec.~\ref{analytical_calculations}, the method and the approximations of
the analytical calculation of the curvature-coupled diffusion coefficient are
presented while sec.~\ref{simulation_methode} introduces the used simulation
scheme. Sec.~\ref{sec:parameters} discusses the choice of parameters used in
both the calculations and the simulations. The presentation of the results in
sec.~\ref{sec:results} is followed by a detailed discussion in
sec.~\ref{sec:discussion}, why simulations lead to smaller diffusion constants
and a possible interpretation of our findings. The paper finishes with some
conclusions and an outlook for future work.

\section{Model}
\label{sec:model}

\subsection{Membrane dynamics}
\label{sec:memb_dyn}

We consider a model membrane in a fluid environment. The membrane is given in
Monge-representation where $\Vektor{r}\equiv(x,y)^T$ is the position in the
$(x,y)$-plane with the deviation $\hrt$ out of this plane. For such
a membrane  Helfrich \cite{helfrich} derived the free energy that to lowest
order has the following form in the Monge-gauge~\cite{safran}
\begin{equation}
  \label{eq:orginal_helfrich}
  \mathcal H_0\left[\hrt\right]=\int\limits_{L^2} \!\dd \Vektor{r}\left[\frac{\kappa}{2}\left(\nabzr\hrt\right)^2+\frac{\sigma}{2}\left(\naber\hrt\right)^2\right],
\end{equation}
with the bending rigidity $\kappa$, the effective surface tension $\sigma$ and
the area $L^2$ in the $\left(x,y\right)$-plane. 
The dynamics of a membrane is given by \cite{seifert}
\begin{equation}
  \label{eq:unperturbed_membrane}
  \partial_t\hrt=-\int\limits_{L^2}\dd\Vektor{r}'\Onsager{r}{r}\frac{\delta \mathcal H_0}{\delta\hrst}+\smvrt
\end{equation}
with the Onsager coefficient $\Onsager{r}{r}$ that takes into account
hydrodynamic interactions with the fluid background. Using the
Fourier-transformation
\begin{eqnarray}
  \hkt&=&\int\limits_{L^2}\dd\Vektor{r}\,\hrt\Exp{-i\Vektor{r}\cdot\Vektor{k}}\\
  h(\mathbf{r},t)&=&\frac{1}{L^2}\sum_{\mathbf{k}}h(\mathbf{k},t)\Exp{i\Vektor{r}\cdot\Vektor{k}}\label{eq:inverse_Fourier}
\end{eqnarray}
one obtains (\ref{eq:unperturbed_membrane}) in Fourier-space
\begin{equation}
  \label{eq:unperturbed_membrane_k}
  \partial_t \hkt=-\Onsagerk{k}E(k)\hkt+\smvkt
\end{equation}
with $E(k)\equiv\kappa k^4+\sigma k^2$ and
$\Onsagerk{k}\equiv1/\left(4\eta k\right)$ which is the Fourier-transformed Onsager
coefficient for a free membrane in a fluid with viscosity
$\eta$~\cite{seifert}. The stochastic force $\smvkt$ obeys the fluctuation-dissipation
theorem
\begin{eqnarray}
  \Mittel{\smvkt}&=&0,\\
  \Mittel{\smvkt\xi^*(\mathbf{k}',t')}&=&2\Onsagerk{k}\frac{L^2}{\beta}\delta\left(t-t'\right)\delta_{\Vektor{k},\Vektor{k}'},\label{eq:fluct_diss}
\end{eqnarray}
where $\beta\equiv(k_bT)^{-1}$ is the inverse temperature. Later on, we need the correlations of $\hkt$ which are given by
\begin{eqnarray}
  \Mittel{\hkt}&=&0,\\
  \label{eq:hkthksts}
  \Mittel{\hkt h^*\!(\Vektor{k}'\!,t')}\! \!& = \!&\!\frac{L^2}{\beta E(k)}\exp[-\Onsagerk{k}E(k)|t\!-\!t'|]\delta_{\Vektor{k},\Vektor{k}'}\!.
\end{eqnarray}
\subsection{Diffusion}
\label{sec:diffusion}
Now we place an inclusion into the membrane that diffuses freely along the
membrane. The dynamics of the inclusion may be described by a
Fokker-Planck-equation (FP-eq.). However, since the diffusive motion takes
place on a curved surface the Laplace-operator needs to be replaced by the
Laplace-Beltrami-operator. This leads to a new
FP-eq. \cite{ellen_projection,ellen_drift}
\begin{equation}
  \partial_tP(\Vektor{r},t)=D\sum\limits_{i,j}\partial_i\sqrt{g}g^{ij}\partial_j \frac{1}{\sqrt{g}}P\left(\Vektor{r},t\right),
\end{equation}
with the diffusion coefficient $D$, the metric $g$ and the inverse metric
tensor $g^{ij}$. In the Monge gauge the metric has the form
$g\equiv1+h_x^2+h_y^2$, while the inverse metric tensor is
\begin{equation}
  g^{ij}\equiv\left(\begin{array}{cc}
    1+h_y^2&-h_xh_y\\-h_xh_y&1+h_x^2
  \end{array}
  \right).
\end{equation}
The subscripts denote partial derivatives, e.g.~$h_x\equiv\partial h/\partial
x$.  The probability $P(\Vektor{r},t)$ of finding the projection
of the inclusion at a position $\Vektor{r}$ is normalized to
$\int\!\dd\Vektor{r}P(\Vektor{r},t)=1$. In the simulations we make
use of a Langevin-equation to describe the motion of the projected particle
position
$\Vektor{R}(t)=\left(X(t),Y(t)\right)^T$. Using
the above FP-eq.~a projected Langevin-eq.~is derived within the Stratonovich
calculus \cite{risken,Haenggi:1982}:
\begin{eqnarray}
  \label{eq:FP_projected}
  \partial_t X(t)&=&D\frac{1}{g(\sqrt{g}+1)}\left(h_Yh_{XY}-h_Xh_{YY}\right)\nonumber\\
  &&+\sqrt{D}\frac{1}{g-1}\left[\left(\frac{h_X^2}{\sqrt{g}}+h_Y^2\right)\stexyt{X}+\right.\nonumber\\&&\qquad\qquad\qquad\left.h_Xh_Y\left(\frac{1}{\sqrt{g}}-1\right)\stexyt{Y}\right]\nonumber,\\
  \partial_t Y(t)&=&D\frac{1}{g(\sqrt{g}+1)}\left(h_Xh_{XY}-h_Yh_{XX}\right)\\
  &&+\sqrt{D}\frac{1}{g-1}\left[h_Xh_Y\left(\frac{1}{\sqrt{g}}-1\right)\stexyt{X}+\right.\nonumber\\&&\qquad\qquad\qquad\left.\left(\frac{h_Y^2}{\sqrt{g}}+h_X^2\right)\stexyt{Y}\right]\nonumber.
\end{eqnarray}
The upper case subscripts express that the partial derivatives at the  particle
position
$\Vektor{R}(t)$ have to be used. 
The stochastic force {\boldmath$\stochteil$}  has zero mean and is delta-correlated:
\begin{eqnarray}
  \Mittel{\stochteil_i}&=&0,\\
  \Mittel{\stochteil_i\left(t\right)\stochteil_j\left(t\right)}&=&2\delta_{ij}\delta\left(t-t'\right).
\end{eqnarray}
Equation (\ref{eq:FP_projected}) comprises a drift that is caused by the
membrane curvature and diffusive terms. The consequences of such a drift term for
a freely diffusing inclusion have been introduced in ref.~\cite{ellen_drift}.
\subsection{Curvature-coupled model}
The equations derived so far apply to a freely diffusing point-like inclusion. If one is
interested in the diffusion of a more realistic inclusion, one has to take the
physical parameters of the inclusion into account. First the inclusion has a
non vanishing area which is set to $\pi a_p^2$. Furthermore, the inclusion has
possibly its own bending rigidity $m$ and maybe a spontaneous curvature
$C_p$. As indicated by ``into the membrane'' the inclusion
completely replaces the membrane at its position. To consider this in the free
energy of the system, one has to add a new energy term for the inclusion and remove the part of the membrane which is
replaced. The additional term caused
by the inclusion leads to the new free
energy
\begin{equation}
  \mathcal H=\mathcal H_0 + \mathcal H_1,
\end{equation}
where 
\begin{multline}
  \label{eq:extention_of_free_energy}
  \mathcal H_1\left[\hrt,\Vektor{R}(t)\right]=\int\limits_{L^2}\!\dd\Vektor{r}\,\grR\times\\\times\left[\frac{m}{2}\left(\nabzr\hrt-C_p\right)^2-\frac{\kappa}{2}\left(\nabzr\hrt\right)^2\right]
\end{multline}
is the correction to Helfrich's free energy $\mathcal H_0$. $\grR$ is a
weighting function for the extension of the particle that we set to be a
Gaussian such that the crossover from particle to membrane is smooth. Taking
into account the area constraint $\int\dd\Vektor{r}\,\grR=\pi a_p^2$ it is
given by
\begin{equation}
  \grR=\Exp{-\frac{\left(\Vektor{r}-\Vektor{R}\right)^2}{a_p^2}}.
\end{equation}
The altered free energy (\ref{eq:extention_of_free_energy}) induces additional forces on the inclusion and the membrane. The membrane dynamics is obtained by replacing $\mathcal H_0$ in (\ref{eq:unperturbed_membrane}) with $\mathcal H$ such that 
\begin{equation}
  \label{eq:membrane_dynamic}
  \partial_t\hrt=-\int\limits_{L^2}\dd \Vektor{r}'\Onsager{r}{r}\left( \frac{\delta\mathcal H_0}{\delta\hrst}+\frac{\delta\mathcal H_1}{\delta\hrst}\right)+\xi(\mathbf{r},t).
\end{equation}
The forces that influence the diffusive behavior of the inclusion can be
calculated by $\Vektor{f}\equiv-\nabeR\mathcal H_1$. Taking into account the
curvature of the membrane in the force term, which needs to be added to the
right hand side of (\ref{eq:FP_projected}), we get the complete equation
of motion for the inclusion
\begin{equation}
  \label{eq:dynamic_inclusion}
  \partial_t\Vektor{R}_i\left(t\right)=\partial_t\Vektor{R}_{proj,i}-\mu\frac{1}{g}\sum_j{g}^{ij}\partial_j\mathcal H_1,
\end{equation}
with the mobility $\mu$ that is related to the intramembrane diffusion
coefficient $D$ via the Einstein relation $D=k_bT\mu$.

With eqs.~\eqref{eq:membrane_dynamic} and \eqref{eq:dynamic_inclusion} the
dynamics of the system is fully determined. Note, that these equations are
coupled since the particle diffusion depends on the shape of the membrane via
the partial derivatives of $h(\mathbf{r},t)$ at the particle position, and the
membrane dynamics on the position of the particle through the additional
energy.

\section{Analytical calculations}
\label{analytical_calculations}
In order to calculate a new curvature-coupling affected diffusion coefficient defined as
\begin{equation}
  \label{eq:dcc_basic}
  D_{cc}\equiv\lim_{t\rightarrow\infty}\frac{\Mittel{\Delta \Vektor{R}^2(t)}}{4t},
\end{equation}
one has to determine the mean square displacement 
\begin{equation}
  \label{eq:mean_square_displacement}
  \Mittel{\Delta \Vektor{R}^2(t)}\equiv\intlim{\dd \tau}{0}{t}\intlim{\dd\tau'}{0}{t}\Mittel{\partial_\tau\Vektor{R}(\tau)\cdot\partial_{\tau'}\Vektor{R}(\tau')},
\end{equation}
by integrating eq.~\eqref{eq:dynamic_inclusion} in time and performing the
thermal average. Since the explicit calculation of the mean square
displacement using the exact equation of motion~\eqref{eq:dynamic_inclusion}
and the full membrane dynamics~\eqref{eq:membrane_dynamic} is not possible
analytically, it is necessary to introduce several approximations.

In order to simplify eq.~\eqref{eq:dynamic_inclusion} we perform a
pre-averaging approximation. This approximation is applicable if for all modes
the membrane relaxation times $(\Lambda(k)E(k))^{-1}$, see
eq.~\eqref{eq:hkthksts}, are considerably shorter than the time $\pi^2/(Dk^2)$
it takes a particle to diffuse the distance given by the corresponding wave
length of the mode. For typical experimental values for bending rigidity
$\kappa$, tension $\sigma$, diffusion coefficients $D$, and system sizes $L$,
this condition is very often fulfilled.  If membrane fluctuations are
``faster'' than the diffusion of the particle it is assumed that the particle
only feels average membrane fluctuations. The applicability of this
approximation for free lateral diffusion is discussed in
ref.~\cite{ellen_drift}. In our current work the pre-averaging approximation
results in the replacement of $\mu\frac{1}{g}g^{ij}$ in
eq.~\eqref{eq:dynamic_inclusion} by $\mu_{proj}\delta_{ij}$ where the
projected free mobility is defined by
$\mu_{proj}\equiv\mu(1+\langle1/g\rangle)/2$~\cite{ellen_projection,ellen_drift}.

We, furthermore, assume that the additional energy $\mathcal{H}_1$ caused by
the insertion of a single particle is small, in order to justify a
perturbation expansion to first order in the particle energy. The consequence
of this approximation is that the dynamics of the membrane is not influenced
by the presence of the inclusion. Thus the membrane dynamics is expressed by
eq.~\eqref{eq:unperturbed_membrane_k}.

Another approximation needs to be employed so as to make analytical
calculations possible. Inserting eq.~\eqref{eq:dynamic_inclusion} into
eq.~\eqref{eq:mean_square_displacement} we see that the mean square
displacement becomes a function of the height correlations $\langle
h(\mathbf{R}(t),t)h(\mathbf{R}(t'),t')\rangle$. These correlations decay with
increasing time difference $|t-t'|$ due to two reasons: during the time
interval the membrane shape changes and the particle position advances. Since
we assume diffusion to be much slower than membrane shape changes we neglect
the effect caused by the particle movement.

Note, that this approximative analytical calculation cannot include a possible
correlation between the particle position and the membrane shape in the
vicinity of the inclusion. In other words, we assume a constant probability for
finding the particle at any point in the system relative to a given membrane
configuration.  This aspect becomes important when we compare with simulation
results, as will be discussed in sec.~\ref{sec:discussion}.

Using the inverse Fourier-transform given in eq.~\eqref{eq:inverse_Fourier}
and applying the previously explained approximations we find for the mean
square displacement
\begin{equation}
  \begin{split}
    &\Mittel{\Delta \Vektor{R}^2\left(t\right)}=4D_{proj}t+\\&
    +m^2C_p^2\mu_{proj}^2\intlim{\dd\tau}{0}{t}\intlim{\dd\tau'}{0}{t}\frac{1}{L^4}\sum_{\Vektor{k}}\sum_{\Vektor{k}'}\left(\pi
    a_p^2\right)^2k^2k'^2\Vektor{k}\cdot\Vektor{k}'\times\\&\times\exp\left\lbrace-i\Vektor{R}\cdot\left(\Vektor{k}+\Vektor{k}'\right)-\frac{\left(k^2+k'^2\right)a_p^2}{4}\right\rbrace\Mittel{\hkt\hksts}
  \end{split}
  \label{eq:dcc_pre_average}
\end{equation}
with the hight correlation function given in eq.~\eqref{eq:hkthksts} and
$D_{proj}=k_bT\mu_{proj}$.

Inserting the resulting
equation for $\Mittel{\Delta \Vektor{R}^2\left(t\right)}$ into
(\ref{eq:dcc_basic}) and performing the long time limit one gets
\begin{multline}
\label{eq:Dcc_sum}
  D_{cc}=D_{proj}+\mu_{proj}^2m^2C_p^2\left(\pi
  a_p^2\right)^2\frac{1}{L^2}\times\\\times\sum_{\Vektor{k}}k^6\Exp{-\frac{k^2a_p^2}{2}}\frac{1}{2\beta
  E^2\left(k\right)\Onsagerk{k}}.
\end{multline}
In this equation it is interesting that there is no need to set a cut-off for
the wavenumber $\Vektor{k}$ since the exponential function $\exp\left\lbrace
-k^2a_p^2/2\right\rbrace$ damps higher $\Vektor{k}$ values. In
ref.~\cite{ellen_projection} a similar calculation for the curvature coupled
diffusion coefficient is performed. There the area function has the form $\pi
a_p^2\delta\left(\Vektor{r}-\Vektor{R}\right)$ and a cut-off is necessary. The
resulting diffusion coefficient agrees with the above diffusion coefficient of
equation \eqref{eq:Dcc_sum} in the limit of vanishing variance of the
Gaussian.

\section{Simulation method}
\label{simulation_methode}

To probe the applicability of our analytical calculations that depend on
several approximations we set up simulations that numerically integrate the
coupled equations of motion for the membrane and the diffusing particle. We
use a square, periodic lattice with $N\times N$ lattice points and the lattice
spacing $\ell$ to map a model membrane with size $L=N\times\ell$, see fig \ref{fig:sketch1}. To simulate
the shape fluctuations of the membrane we numerically integrate the
appropriate equation of motion. In order to reduce the computational effort
and to compare with the analytical calculations introduced in the previous
section we evaluate the unperturbed Langevin-equation
\eqref{eq:unperturbed_membrane_k} discretely in time. 
These calculations are
performed in Fourier-space since the equations of motion for the height
function modes $\hkt$ decouple. Due to the periodic boundary conditions the
wave vectors are of the form $\mathbf{k}=2\pi(l,n)/L$ with $l$ and $n$ being
integers. Since the hight function $h(\mathbf{r})$ is a real function defined
on a $N\times N$ lattice the relation $h(\mathbf{k},t)=h^*(-\mathbf{k},t)$
applies leading to the restriction $-N/2<l,n\leqslant N/2$.

Regarding the fluctuation-dissipation theorem~\eqref{eq:fluct_diss} it is
obvious that fluctuations of $h(\mathbf{k}=0,t)$ would diverge for the 
Onsager-coefficient $\Onsagerk{\Vektor{k}\to 0}\to\infty$. The dynamics of the height function mode
$h(\mathbf{k}=0,t)$ corresponds to the center of mass movement of the whole
membrane. Due to the irrelevance of this movement in the determination of the
lateral diffusion coefficient we set $\Lambda(\mathbf{k}=0)=0$ and keep
$h(\mathbf{k}=0,t)=0$ fixed at all times.

In order to choose an appropriate discrete time step $\Delta t$ that ensures
that numerical errors are small it is necessary to point out that the largest
$\Vektor{k}$-vector possible $k_{max}=\sqrt{2}\pi N/L$ in the given lattice
determines the smallest time scale of the membrane as can be seen in
eq.~\eqref{eq:hkthksts}. The used time step $\Delta t$ in the simulation
should be smaller than this smallest time scale.

The dynamics of the inclusion is given by a discrete version of
eq.~\eqref{eq:dynamic_inclusion} that is also numerically integrated in time.
Since it is coupled to the membrane equation (\ref{eq:unperturbed_membrane_k})
via the derivatives $h_X$, $h_Y$, etc., of the membrane configuration at the
position of the inclusion, the temporal evolution during a discrete time step
$\Delta t$ consists of an update of the membrane shape and the particle
position.  In contrast to the membrane dynamics the motion of the inclusion is
calculated in real space. The required derivatives of $\hrt$ are, therefore,
determined in Fourier-space and then transformed to real space using routines
of the FFTW-libraries~\cite{fftw}. We allow off-lattice diffusion for the
inclusion which is necessary if the average time it takes a particle to
diffuse a distance $\ell$ is much larger than the corresponding membrane
relaxation time. Thus the derivatives at the position of the inclusion are
determined by a distance weighted linear extrapolation of the four nearest
lattice sites. In order to minimize the numerical error caused by not
following the membrane surface correctly the displacement per time step is to
be set to a small fraction of the lattice spacing $\ell$. With the above
explained restriction for the time step in order to describe the membrane
shape evolution with sufficient accuracy there are overall two conditions that
must be fulfilled in the choice of $\Delta t$.

Apart from obvious computational limits the determination of how long
simulation runs should at least be is again dictated by two time scales. On
the one hand the length of the simulation should be several times the longest
relaxation time of the membrane, which is given by the smallest possible wave
vector, to ensure that the membrane shape has passed through an adequate
amount of independent configurations. On the other hand it is preferable that
the inclusion has enough time on average to cover a distance of several
lattice sites.

A more detailed description of the simulation method is given
in ref.~\cite{ellen_drift}, where the corresponding scheme for free particle
diffusion has been introduced.
	
\section{Parameters}
\label{sec:parameters}
Before we present our results we will introduce the used parameters. These are
the same for the analytical calculations and the simulations. As already
mentioned in the description of the simulation method we use a discrete
membrane with a lattice spacing of $\ell$ that we set to
$\ell=10\mathrm{nm}$. This choice reflects a compromise between the wish to simulate
reasonably sized systems and the computational limitation in the number of
lattice sites. A decrease in the lattice spacing for a constant system size,
i.e.~an increase in the number of lattice sites, introduces additional large
$k$-values that contribute only weakly to membrane fluctuations, see
eq.~\eqref{eq:hkthksts}, or the curvature-coupled diffusion
coefficient~\eqref{eq:Dcc_sum}.  
$\ell$ is one of the basic units. The others are the time given in seconds
$\mathrm s$ and the thermal energy that is $\beta^{-1}=4.14\times
10^{-21}\mathrm{J}$ at room temperature. All parameters of the system are
given in units of $\ell$, $\mathrm{s}$, and $\beta$. For the determination of
the membrane parameters we look at typical experiments and extract a range of
$5$ to $50$ for the bending rigidity $\kappa\beta$ and $10^{-9}$ to $10^{-6}
\mathrm{J/m^2}$ for the effective tension $\sigma$ that corresponds to
$10^{-5}$ to $10^{-2}$ for $\sbl$. At around $\sbl=10$ rupture of the membrane
occurs. In the experiments it is, furthermore, common to use water as a
surrounding medium with a viscosity of $\eta=10^{-3}\mathrm{kg/ms}=2.47\times
10^{-7}\mathrm{s}/\left(\beta\ell^3\right)$. The last parameter of the
membrane is the size $L$ that is related to the number of lattice points $N$
in each direction via $L=N\times \ell$. Since a sufficient number of wave
vectors $\Vektor{k}$ are considered for a $50\times50$ lattice we use a system
size of $L=50\ell$ that corresponds to $0.5\mathrm{\mu m}$. For the parameters
of the inclusion we choose $C_p\ell=2$, $m\beta=2\kappa\beta$ and $a_p=1\ell$
as in our previous calculations \cite{ellen_projection}. The bare diffusion
coefficient $D$, however, has to be chosen carefully. Since the analytical
calculations rely on a pre-averaging approximation we have to set $D$
sufficiently small. Therefore, the time scale of the diffusion
$\tau_{D}=\pi^2/\left(Dk_{min}^2\right)$ must be much longer than the largest
membrane time. This time is given by $\tau_{memb,max}=4\eta/\left(\kappa
k^3_{min}+\sigma k_{min}\right)$ with the absolute value of the smallest wave
vector $k_{min}=2\pi/L$ which corresponds to the longest length $L$ in the
system. The comparison of the two time scales leads to $D\ll\pi^2\left(\kappa
k_{min}+\sigma k_{min}^{-1}\right)/\left(4\eta\right)\sim 10^6\ell^2/s$. We
choose $D=5\times10^4\ell^2/s$. This choice is good for all values of $\kb$
and $\sbl$ in the selected range. For the total length of the simulations one
has to keep in mind that it has to be several times longer than the time scale
of the longest wave vector. The time step $\Delta t$ however has to be so
small that it is smaller than the shortest membrane time $\tau_{memb,min}$ and
that the inclusion diffuses only a short distance. We set the total length to
$1\mathrm{ms}$ and $\Delta t =10^{-9}\mathrm s$ such that each simulation run
comprises $10^6$ time steps.  This choice of the time step $\Delta t$ is
applicable for small bending rigidities $\kb$.

Note, that an increase of $\kb$ and $\sbl$ leads to smaller
time scales $\tau_{memb}$ of the membrane. In order not to increase computing
time we keep $\Delta t$ fixed for all considered $\kb$ and $\sbl$. This,
however, will lead to a slight increase in numerical errors for the modes
$\hkt$ with very large wave vectors in membranes with large $\kb$ and
$\sbl$. Since fluctuations of these modes, see eq.~\eqref{eq:hkthksts},
are rather small these errors are negligible in the determination of
$D_{cc}$.

\section{Results}
\label{sec:results}
\begin{figure}
 \centering
 \includegraphics{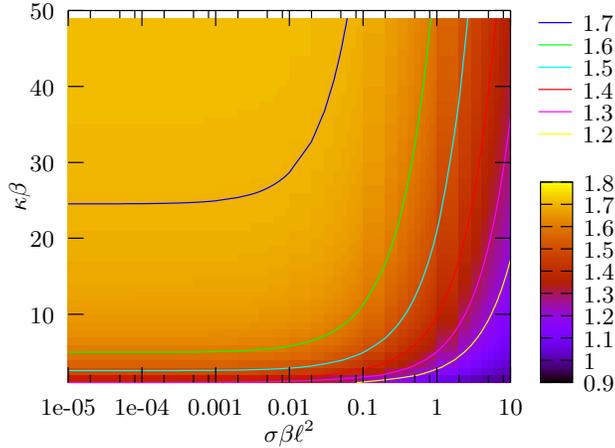}
 \caption{Ratio of the curvature-coupled diffusion coefficient to the free
   diffusion $D_{cc}/D$ as a function of the bending rigidity $\kb$ and the
   effective tension $\sbl$.
  \label{fig:numeric_summation}}
 \end{figure}
The analytically derived curvature-coupled diffusion coefficient $D_{cc}$ is
calculated by numerical summation of equation \eqref{eq:Dcc_sum} using
\textit{Mathematica} for the whole range of parameters given in the previous
section. The ratio $D_{cc}/D$ of the curvature-coupled $D_{cc}$ and
the free intramembrane diffusion coefficient $D$ is plotted as a function of
the bending rigidity $\kb$ and the effective surface tension $\sbl$ in
fig.~\ref{fig:numeric_summation}. The ratio increases for brighter
colors. In the figure one can see that for small $\sbl$ the ratio increases
very strongly for $\kb<5$. With a further increase of the bending rigidity $\kb$ the ratio reaches a
plateau. Here the strengthening of the forces $\mathbf f$ caused by an
increasing bending rigidity of the inclusion $m\beta$ is compensated by the
fact that thermal fluctuations become weaker for increasing $\kb$. The
increase of $\sbl$ by about three orders of magnitude leads to no significant
effect. However, for even larger values of $\sbl$ the ratio decreases fast to
one for small $\kb$. This happens since a large surface tension $\sbl$ also
damps the thermal fluctuations of the membrane and, therefore, the additional
force on the inclusion is small. Increasing $\kb$ the ratio also increases for large $\sbl$ but does not reach
the same hight as for small tensions. In this case the increase of $D_{cc}/D$
for high values of $\kb$ is also compensated by the damping caused by the
surface tension. 

Overall, our calculations show that the curvature-coupling, which leads to an
additional force $\mathbf f$ on the inclusion, enhances the inclusion's
diffusion rate in the investigated parameter range. It is noteworthy that the
ratio is always bigger than one despite the fact that the projection alone
would lead to a ratio smaller than one \cite{ellen_projection}.

Results from variations of the other parameters, like $L$, $C_p$, etc., are
not plotted but the effect on $D_{cc}$ can easily be obtained from
eq.~\eqref{eq:Dcc_sum}. However, one has to keep in mind that the effect of
the inclusion on the membrane has to be small in order for the perturbation
theory to be applicable.
\begin{figure}
 \centering
 \includegraphics{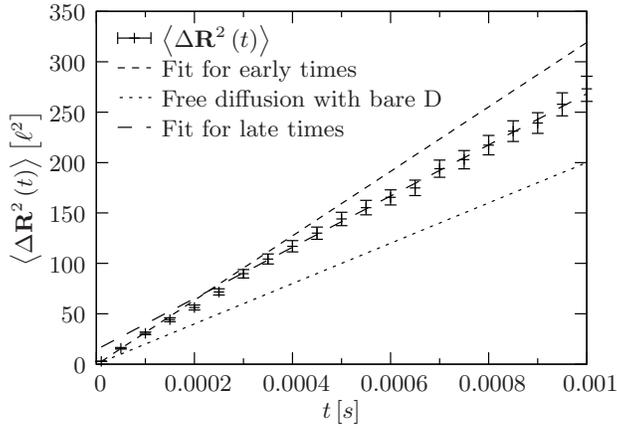}
 \caption{Mean square displacement determined from simulation data as a
   function of time for $\kb=5$ and $\sbl=0$.  Plotted with the data points
   are the two fits for the diffusion coefficient at the beginning and the end
   of the simulations. Furthermore, the expected mean square displacement for
   bare free diffusion is displayed.
 \label{fig:example_diffusion}}
\end{figure}
\begin{figure}
 \centering
 \includegraphics{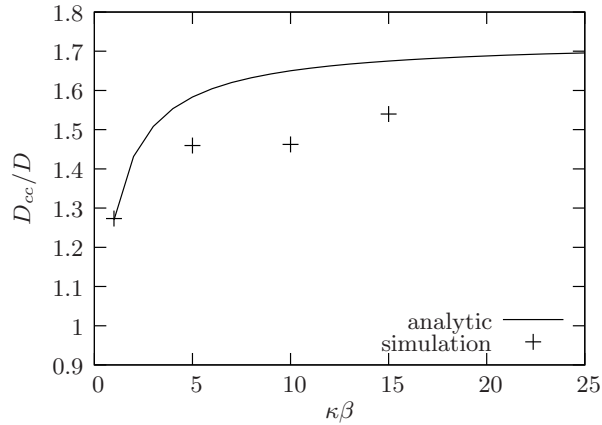}
 \caption{Ratio $D_{cc}/D$ of the curvature-coupled to the intramembrane diffusion coefficient as a function of bending
 rigidity $\kappa\beta$ for fixed surface tension $\sbl=1\times10^{-2}$. Most
 of the simulated data points  are smaller than the analytical curve
 determined by eq.~\eqref{eq:Dcc_sum}.
 \label{fig:sigma-2}}
\end{figure}
\begin{figure}
 \centering
 \includegraphics{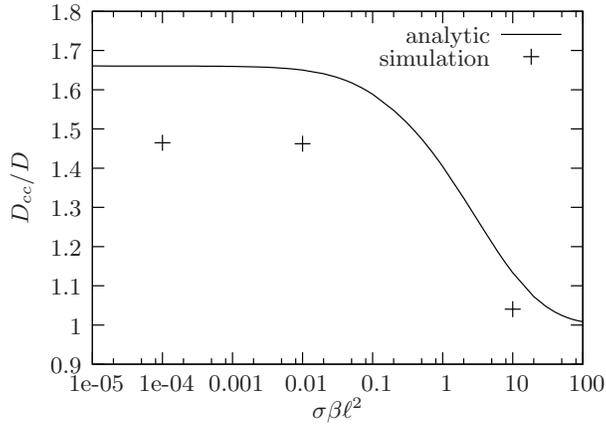}
 \caption{Ratio $D_{cc}/D$ as a function of $\sbl$ for the fixed bending rigidity
 $\kb=10$. The simulated values (+) follow
 the analytical curve (line) of eq.~\eqref{eq:Dcc_sum} only qualitatively. 
 \label{fig:kappa10}}
\end{figure}

In the simulations, we also use the unperturbed membrane equation of motion
(\ref{eq:unperturbed_membrane_k}) and the same parameter sets as for the
numerical summation just discussed. As we are interested in results in thermal
equilibrium each membrane starts in a random configuration and has $1$ ms to
equilibrate before the inclusion is placed in its center. $1$ ms is about five
times the longest membrane relaxation time so we can be sure that the membrane
is in thermal equilibrium. Since we obtain only one particle trajectory per independent simulation run we
average over 500 simulations with the same set of parameters to get the mean
square displacement $\Mittel{\Delta \Vektor{R}^2(t) }$. An example for
$\Mittel{\Delta \Vektor{R}^2(t)}$ as a function of time is plotted in figure
\ref{fig:example_diffusion}. The slope of the resulting straight line at late
times corresponds to $D_{cc}$, see eq.~\eqref{eq:dcc_basic}.  The determined
values for $D_{cc}/D$ from the simulations (+) are plotted with the analytical curve
(line) in fig.~\ref{fig:sigma-2} as a function of $\kb$ for a constant $\sbl$
and in fig.~\ref{fig:kappa10} as a function of $\sbl$ for a constant bending
rigidity $\kb$. We see that the simulations follow qualitatively the
analytical curve but the values are about $10\%$ smaller than expected.

\section{Discussion}
\label{sec:discussion}
Both of our approaches demonstrate that curvature-coupling enhances
diffusion. This result is plausible for the following reason. Due to the
membrane fluctuations the positions that are favorable for the diffusing
particle are constantly changing. Thus the particle is subject to changing
forces leading to enhanced movement of the particle, which in turn leads to a
higher diffusion coefficient.  The resulting enhanced diffusion coefficient is
caused by forces, which are still thermal. Therefore, the system is still in
equilibrium and the fluctuation-dissipation-theorem is applicable, such that
an increased effective mobility or a reduced effective friction of the
particle can be determined.

On the quantitative side, the analysis of the simulation data reveals a
diffusion coefficient that is about $10\%$ smaller than we expect from the
analytical calculations. In these calculations several approximations that we
have explained in sec.~\ref{analytical_calculations} are applied to calculate
the mean square displacement $\Mittel{\Delta \Vektor{R}^2(\tau)}$
(\ref{eq:mean_square_displacement}). The dominant contribution to $D_{cc}$ is
$\Mittel{\Vektor{f}(\tau,\Vektor{R}(\tau))\cdot\Vektor{f}(\tau+\Delta\tau,\Vektor{R}(\tau+\Delta\tau))}$
with the force
$\Vektor{f}\left(\tau,\Vektor{R}\left(\tau\right)\right)\equiv-\nabeR\mathcal
H_1[h(\mathbf{r},\tau),\mathbf{R}(\tau)]$, see
eqs.~\eqref{eq:dynamic_inclusion}, \eqref{eq:mean_square_displacement}. Hence
we investigate this force correlation function.
\begin{figure}
 \centering
 \includegraphics{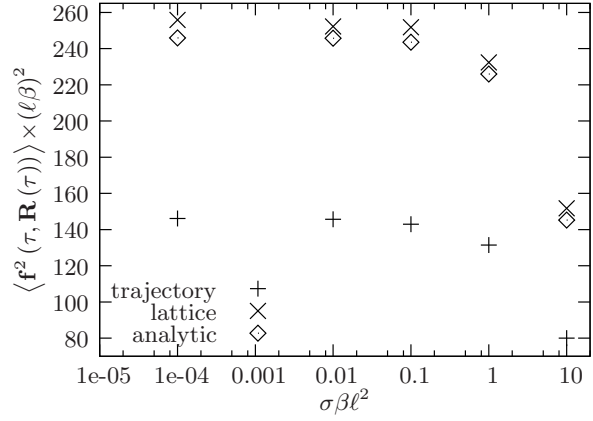}
 \caption{Average over the squared force on the inclusion for equal times
 $\Mittel{\Vektor{f}^2\left(\tau,\Vektor{R}\left(\tau\right)\right)}$ as a function of the effective tension
 $\sbl$ for the fixed bending rigidity $\kb=5$. The figure shows a good agreement between the analytical points ($\diamondsuit$) and those achieved by averaging over a fixed point on the lattice (x). The
 averaging along the trajectory (+) leads to smaller values.
 \label{fig:strength_kappa5}}
\end{figure}
\begin{figure}
 \centering
 \includegraphics{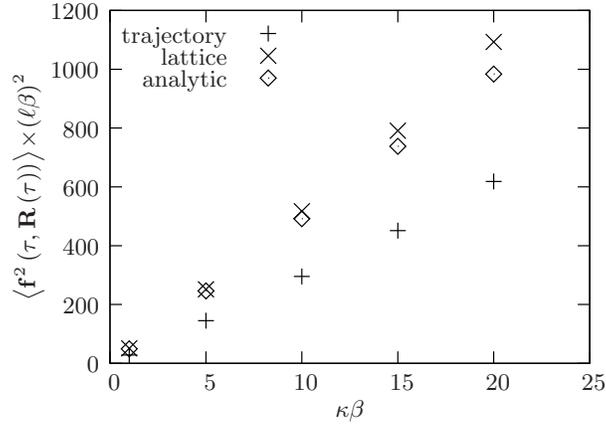}
 \caption{Average over the squared force on the inclusion for equal times
 $\Mittel{\Vektor{f}^2(\tau,\Vektor{R}(\tau))}$ as a function of the bending
 rigidity $\kb$ for the fixed surface tension $\sbl=0$. The figure shows a
 good agreement between the analytical points ($\diamondsuit$) and those achieved by
 averaging over a fixed point on the lattice (x). The averaging along the
 trajectory (+) leads to smaller values.
 \label{fig:strength_sigma0}}
\end{figure}

First we consider the averaged quadratic force for $\Delta\tau=0$ along the
trajectory $\Vektor{R}(\tau)$ of the inclusion to see whether differences in
the strength of the correlations occur. Then we will regard the dependence of
the correlations on the time interval $\Delta\tau$. To examine the strength of
the correlations for $\Delta\tau=0$ the quadratic force acting on the
inclusion is determined at each time step of the simulation and then averaged
over all values. For several sets of parameters we receive the values that are
plotted in fig.~\ref{fig:strength_kappa5} as a function of $\sbl$ for a
constant $\kb$ and in fig.~\ref{fig:strength_sigma0} as a function of $\kb$
for a constant $\sbl$. The values resulting from the analytical calculations
are also plotted in these figures. We observe a difference between analytical
and simulation results that is on the same order of magnitude as the
difference in the diffusion coefficients. In the analytical calculations we
assume that the probability of finding the inclusion at a particular position
is the same for any point of the membrane. If we calculate, using the
simulation data, the mean square of the force the inclusion would be exposed
to if it were fixed at some arbitrary position on the membrane we obtain
values very close to the mean squared force resulting from the analytical
calculations. These values are also plotted in figs.~\ref{fig:strength_kappa5}
and \ref{fig:strength_sigma0}. The differences in
fig.~\ref{fig:strength_kappa5} between the analytical calculations and the
average for a fixed point are caused by the time step $\Delta t$ of the
simulations that induces bigger numerical errors for higher values of $\kb$ as
previously explained in section \ref{sec:parameters}. Overall, averaging over
the whole lattice in the calculations leads to seemingly higher forces than
along the actual particle trajectory. A possible explanation for this reduced
force is that most of the time the inclusion is close to a local minimum of
the free energy. The extrema of the energy are created by the membrane shape,
which is constantly changing due to thermal fluctuations. Therefore, the
positions of the extrema will also move along the membrane. As the negative
gradient of the energy is always pointing towards the nearest local minimum
the inclusion will predominantly move in the direction of the nearest local
minimum. For a fast enough diffusion rate the inclusion is capable of
following a local minimum.

Due to the fact that for each simulation run a new thermally equilibrated
membrane is used and the particle is always placed in the center of the
membrane, it is very unlikely that the inclusion is initially close to an
energy minimum. Therefore, the inclusion is exposed to higher forces at
the beginning than at later times. Since higher forces go along with a higher
diffusion rate we expect to observe two diffusion coefficients from the
simulation data: a smaller one for late and a larger one for early
times. Indeed such a behavior occurs as one can see in
fig.~\ref{fig:example_diffusion} where the linear fits to the mean square
displacement for early and late times are plotted. In this example the
crossover is at about $0.2$ms. Comparing the resulting diffusion coefficients
with the analytical values, the one determined at the beginning of the
simulation agrees reasonably well with the analytically determined diffusion
coefficient. The inclusion starts with a higher diffusion rate and then, after
a variable time period, finds a local minimum, which it tries to follow.

\begin{figure}
 \centering
 \includegraphics{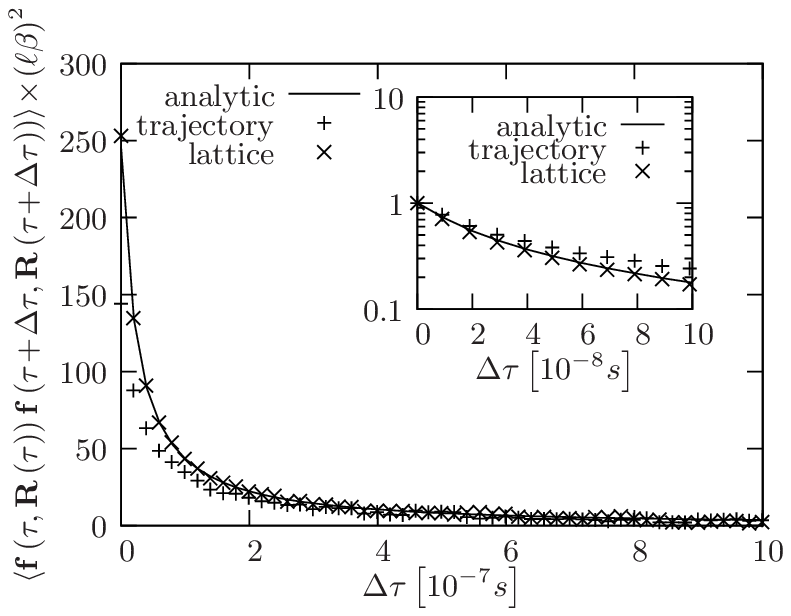}
 \caption{Time correlation function of the additional force on the inclusion
 $\Mittel{\Vektor{f}(\tau,\Vektor{R}(\tau))\cdot\Vektor{f}(\tau+\Delta\tau,\Vektor{R}(\tau+\Delta\tau))}$
 for the bending rigidity $\kb=5$ and the effective tension $\sbl=0$ as a
 function of the time difference $\Delta\tau$. In the inset the first
 $10^{-7}s$ of the time correlation function are plotted scaled to one for
 $\Delta \tau=0$.  A good agreement is found for the analytical curve (line)
 and the average over five fixed lattice points (x). The time dependence of the
 force correlations along the trajectory (+) is different from the analytical result.}
 \label{fig:time_correlation}
\end{figure}

Now that we have found that the force acting on the inclusion is reduced in
the simulations we consider the time correlations of the force
$\Mittel{\Vektor{f}(\tau,\Vektor{R}(\tau))\cdot\Vektor{f}(\tau+\Delta\tau,\Vektor{R}(\tau+\Delta\tau))}$. We
expect to see a reduction of the correlations caused by the inclusion
following an energy minimum but in addition the time dependency will be
influenced by the movement of the particle position
$\Vektor{R}(\tau)$. The obtained correlation function along the
trajectory is plotted in fig.~\ref{fig:time_correlation} together with the
analytical result for $\kb=5$ and vanishing tension. For a better
representation the functions are normalised to one and plotted for small
$\Delta \tau$ in the inset of the figure. It is obvious that in addition to
the altered start values the time dependence is also different. The decay of
the correlations along the trajectory is slower than for the analytical
curve. To demonstrate that this altered decay is caused by the motion of the
inclusion we choose five fixed points of the lattice and determine, from the
simulation data, the time
correlation function of the force that would act on the inclusion if it were
fixed at these points. The average over these points  is also plotted in
fig.~\ref{fig:time_correlation} and agrees well with the analytical
curve. The altered decay along the trajectory is an effect caused by the
motion of the particle and indicates that the force correlations are stronger
along the trajectory. This fact does not only lead to a slightly higher diffusion rate
but also shows that the forces for a series of time steps point in similar
directions which corroborates our interpretation that the inclusion tries to
follow a local minimum. Although a larger decay time of correlations leads to
enhanced diffusion the observed diffusion rate is smaller than the
analytically calculated value since the effect that forces close to energy minima
are reduced dominates.

\section{Conclusions}
\label{sec:conclusions}
In this paper, we have derived a model for the interaction of an inclusion with a
model membrane and have investigated the influence of this interaction on the
diffusion of the inclusion. In the model the inclusion is a physical object
with an area, a spontaneous curvature and a bending rigidity. In analogy to
Helfrich's free energy a new free energy is derived and with this coupled
equations of motion for the inclusion and the membrane dynamics. Using these
stochastic equations and employing several approximations we calculate the
curvature-coupled diffusion coefficient $D_{cc}$. To assess the quality of the
approximations in this analytical calculation that assumes a weak perturbation
of the membrane we set up simulations. These simulations that numerically
integrate the coupled equations of motion for the membrane and the inclusion
are also based on the unperturbed membrane equation.

Both, the simulations and our analytical approach, clearly display that the
additional force on the inclusion caused by the interaction between particle
and membrane leads to a significant increase in the diffusion coefficient
compared to bare intramembrane diffusion. However,
comparing the analytical calculations with the simulation results one finds
that the curvature-coupled diffusion coefficient in the simulations is about
$10\%$ smaller than analytically expected, but follows the behavior
qualitatively. A closer look at the forces on the inclusion shows that the
averaged forces along the trajectory of the inclusion are smaller than the
averaged notional forces acting at any arbitrary fixed point of the
lattice. The average over the latter forces, however, has a good agreement
with the values from the analytical calculations. Since smaller forces
correspond to local extrema of the free energy, and the gradient, i.e.~the
force, always points to the nearest local minimum it is likely that the
inclusion is in the vicinity of such a local minimum of the energy most of the
time. If the mobility of the inclusion is big enough the inclusion is able to
follow such a local minimum. Another point for this interpretation is that we
place the inclusion in a new thermally equilibrated membrane for each
simulation run. Hence, the inclusion is not necessarily close to a local
minimum at the beginning and the diffusion rate should be faster for early
times of the simulations. Considering the simulation data we find indeed that
the diffusion for early times is about the same as the diffusion rate expected
from the analytical calculation. After a short time the diffusion rate
decreases and then remains constant. This corroborates the assumption that the
inclusion meets a local minimum after some time and then tries to follow it.

Our study leads to the conclusion that the analytical calculations provide
qualitative values for the curvature-coupled diffusion coefficient for a given
set of parameters. For a quantitative value of the curvature-coupled diffusion
coefficient simulations are necessary that take the effects of the movement of
the inclusion into account.

In this paper we use the unperturbed membrane equation of motion. In ongoing
work we are investigating the influence of the inclusion on the membrane
dynamics and the diffusion by use of simulations that incorporate the perturbed
membrane equation of motion. Then, the particle does not only adapt to
the membrane but the membrane also adjusts to the particle. This will possibly
lead to a further reduction of the diffusion coefficient compared to the
analytical calculation.  By comparing these results with those of the present
paper it will be possible to determine the parameter range in which the
perturbation may be neglected. We intend to also study the diffusion of
several inclusions in a membrane in order to investigate possible cluster
formation induced by membrane mediated interactions between the
inclusions. Aside from this one may also be interested in other forms of
interactions between the membrane and the inclusion or additional interactions
between the inclusions.

The investigation of several possible interactions and the resulting effects
on the diffusion coefficient of inclusions as a function of membrane
parameters may help to understand experimental data better and should finally
lead to a deeper insight of the diffusion processes in biological membranes.


\begin{thebibliography}{10}

\bibitem{Lippincott:2001}
Lippincott-Schwartz,~J.;\ \ Snapp,~E.;\ \ Kenworthy,~A.
  \textit{Nat.~Rev.~Mol.~Cell Bio.} \textbf{2001,} \textsl{2,} 444-456.

\bibitem{Marguet:2006}
Marguet,~D.;\ \ Lenne,~P.~F.;\ \ Rigneault,~H.;\ \ He,~H.-T. \textit{EMBO J.}
  \textbf{2006,} \textsl{25,} 3446-3457.

\bibitem{Lommerse:2004}
Lommerse,~P.~H.~M.;\ \ Spaink,~H.~P.;\ \ Schmidt,~T. \textit{BBA-Biomembranes}
  \textbf{2004,} \textsl{1664,} 119-131.

\bibitem{Chen:2006}
Chen,~Y.;\ \ Lagerholm,~B.~C.;\ \ Yang,~B.;\ \ Jacobson,~K. \textit{Methods}
  \textbf{2006,} \textsl{39,} 147-153.

\bibitem{Reits:2001}
Reits,~E.~A.~J.;\ \ Neefjes,~J.~J. \textit{Nat.~Cell Biol.} \textbf{2001,}
  \textsl{3,} E145-E147.

\bibitem{Lippincott:2003}
Lippincott-Schwartz,~J.;\ \ Altan-Bonnet,~N.;\ \ Patterson,~G.~H.
  \textit{Nat.~Cell Biol.} \textbf{2003,} \textsl{5,} S7-S14.

\bibitem{kohl}
Kohl,~T.;\ \ Schwille,~P. \textit{Adv.~Biochem.~Eng.~Biot.} \textbf{2005,}
  \textsl{95,} 107-142.

\bibitem{Thompson:2002}
Thompson,~N.~L.;\ \ Lieto,~A.~M.;\ \ Allen,~N.~W.
  \textit{Curr.~Opin.~Struct.~Biol.} \textbf{2002,} \textsl{12,} 634-641.

\bibitem{saxton}
Saxton,~M. \textit{Annu. Rev. Biophys. Biomol. Struct.} \textbf{1997,}
  \textsl{26,} 373-399.

\bibitem{Kusumi:2005}
Kusumi,~A.;\ \ Nakada,~C.;\ \ Ritchie,~K.;\ \ Murase,~K.;\ \ Suzuki,~K.;\ \
  Murakoshi,~H.;\ \ Kasai,~R.~S.;\ \ Kondo,~J.;\ \ Fujiwara,~T.
  \textit{Annu.~Rev.~Biophys.~Biomol.~Struct.} \textbf{2005,} \textsl{34,}
  351-378.

\bibitem{Tomishige:1998}
Tomishige,~M.;\ \ Sako,~Y.;\ \ Kusumi,~A. \textit{J. Cell Biol.} \textbf{1998,}
  \textsl{142,} 989-1000.

\bibitem{Edidin:1991}
Edidin,~M.;\ \ Kuo,~S.~C.;\ \ Sheetz,~M.~P. \textit{Science} \textbf{1991,}
  \textsl{254,} 1379-1382.

\bibitem{Sbalzarini:2006}
Sbalzarini,~I.~F.;\ \ Hayer,~A.;\ \ Helenius,~A.;\ \ Koumoutsakos,~P.
  \textit{Biophys.~J.} \textbf{2006,} \textsl{90,} 878-885.

\bibitem{Aizenbud:1982}
Aizenbud,~B.~M.;\ \ Gershon,~N.~D. \textit{Biophys. J.} \textbf{1982,}
  \textsl{38,} 287-293.

\bibitem{Holyst:1999}
Ho{\l}yst,~R.;\ \ Plewczy\'{n}ski,~D.;\ \ Aksimentiev,~A. \textit{Phys.~Rev.~E}
  \textbf{1999,} \textsl{60,} 302-307.

\bibitem{Plewczynski:2000}
Plewczy\'{n}ski,~D.;\ \ Ho{\l}yst,~R. \textit{J.~Chem.~Phys.} \textbf{2000,}
  \textsl{113,} 9920-9929.

\bibitem{Faraudo:2002}
Faraudo,~J. \textit{J.~Chem.~Phys.} \textbf{2002,} \textsl{116,} 5831-5841.

\bibitem{King:2004}
King,~M.~R. \textit{J.~Theor.~Biol.} \textbf{2007,} \textsl{227,} 323-326.

\bibitem{Yoshigaki:2007}
Yoshigaki,~T. \textit{Phys.~Rev.~E} \textbf{2007,} \textsl{75,} 041901.

\bibitem{Naji:2007}
Naji,~A.;\ \ Brown,~F.~L.~H. \textit{J.~Chem.~Phys.} \textbf{2007,}
  \textsl{126,} 235103.

\bibitem{Brochard:1975}
Brochard,~F.;\ \ Lennon,~J.~F. \textit{J. Phys. (Paris)} \textbf{1975,}
  \textsl{11,} 1035-1047.

\bibitem{seifert}
Seifert,~U. \textit{Adv. in Phys.} \textbf{1997,} \textsl{46,} 13-137.

\bibitem{Gustafsson:1997}
Gustafsson,~S.;\ \ Halle,~B. \textit{J.~Chem.~Phys.} \textbf{1997,}
  \textsl{106,} 1880-1887.

\bibitem{ellen_projection}
Reister,~E.;\ \ Seifert,~U. \textit{Europhys. Lett.} \textbf{2005,}
  \textsl{71,} 859-865.

\bibitem{Gov:2006}
Gov,~N.~S. \textit{Phys.~Rev.~E} \textbf{2006,} \textsl{73,} 041918.

\bibitem{ellen_drift}
Reister-Gottfried,~E.;\ \ Leitenberger,~S.;\ \ Seifert,~U. \textit{Phys. Rev.
  E} \textbf{2007,} \textsl{75,} 011908.

\bibitem{kahya}
Kahya,~N.;\ \ Wiersma,~D.;\ \ Poolman,~B.;\ \ Hoekstra,~D.
  \textit{J.~Biol.~Chem.} \textbf{2002,} \textsl{277,} 39304-39311.

\bibitem{Vereb:2003}
Vereb,~G.;\ \ \mbox{Sz\"oll\H{o}si},~J.;\ \ Matk\'{o},~J.;\ \ Nagy,~P.;\ \
  Farkas,~T.;\ \ V\'{\i}gh,~L.;\ \ M\'{a}tyus,~L.;\ \ Waldmann,~T.~A.;\ \
  Damjanovich,~S. \textit{Proc.~Natl.~Acad.~Sci.~USA} \textbf{2003,}
  \textsl{100,} 8053-8058.

\bibitem{Forstner:2006}
Forstner,~M.;\ \ Yee,~C.;\ \ Parikh,~A.;\ \ Groves,~J.
  \textit{J.~Am.~Chem.~Soc.} \textbf{2006,} \textsl{128,} 15221-15227.

\bibitem{Weikl:1998}
Weikl,~T.~R.;\ \ Kozlov,~M.~M.;\ \ Helfrich,~W. \textit{Phys. Rev. E}
  \textbf{1998,} \textsl{57,} 6988--6995.

\bibitem{Weikl:2001}
Weikl,~T.~R. \textit{Europhys.~Lett.} \textbf{2001,} \textsl{54,} 547-553.

\bibitem{Fournier:1999}
Dommersnes,~P.~G.;\ \ Fournier,~J.-B. \textit{Eur.~Phys.~J.~B} \textbf{1999,}
  \textsl{12,} 9-12.

\bibitem{Fournier:2003}
Dommersnes,~P.~G.;\ \ Fournier,~J.-B. \textit{Eur.~Phys.~J.~E} \textbf{2003,}
  \textsl{11,} 141-146.

\bibitem{Fosnaric:2006}
Fo\u{s}nari\u{c},~M.;\ \ Igli\u{c},~A.;\ \ May,~S. \textit{Phys.~Rev.~E}
  \textbf{2006,} \textsl{74,} 051503.

\bibitem{Cooke:2006}
Cooke,~I.~R.;\ \ Deserno,~M. \textit{Biophys. J.} \textbf{2006,} \textsl{91,}
  487-495.

\bibitem{Deserno:2007}
Reynwar,~B.~J.;\ \ Illya,~G.;\ \ Harmandaris,~V.~A.;\ \ M\"uller,~M.~M.;\ \
  Kremer,~K.;\ \ Deserno,~M. \textit{Nature} \textbf{2007,} \textsl{447,}
  461-464.

\bibitem{Blood:2006}
Blood,~P.~D.;\ \ Voth,~G.~A. \textit{Proc.~Natl.~Acad.~Sci.~USA} \textbf{2006,}
  \textsl{103,} 15068-15072.

\bibitem{Ayton:2007}
Ayton,~G.~S.;\ \ Blood,~P.~D.;\ \ Voth,~G.~A. \textit{Biophys. J.}
  \textbf{2007,} \textsl{92,} 3595-3602.

\bibitem{helfrich}
Helfrich,~W. \textit{Z.~Naturforsch.~C} \textbf{1973,} \textsl{28,} 693-703.

\bibitem{safran}
Safran,~S. \textit{Statistical Thermodynamics of Surfaces, Interfaces, and
  Membranes;} Addison-Wesley Publishing Company: Cambridge, 1994.

\bibitem{risken}
Risken,~H. \textit{The Fokker-Planck Equation;} Springer Verlag: Heidelberg,
  2nd ed.; 1989.

\bibitem{Haenggi:1982}
H\"anggi,~P.;\ \ Thomas,~H. \textit{Phys.~Rep.} \textbf{1982,} \textsl{88,}
  207-319.

\bibitem{fftw}
Frigo,~M.;\ \ Johnson,~S.~G. ``FFTW - for version 3.1.2'',  www.fftw.org, 2006.


\end{thebibliography}
\providecommand{\refin}[1]{\\ \textbf{Referenced in:} #1}

\end{document}